\documentclass[prl,twocolumn,nofootinbib,letter,superscriptaddress]{revtex4-1}
\usepackage{graphicx,psfrag}
\usepackage{bm}
\usepackage{mathbbol,verbatim}
\usepackage{slashed}
\usepackage{graphics}
\usepackage{color,ulem, amsmath, amssymb}
\usepackage{enumitem}  
\definecolor{greeen}{rgb}{0.03,0.84,0.13}
\definecolor{test}{rgb}{0.03,0.74,0.33}
\definecolor{viol}{rgb}{0.44,0,0.94}
\definecolor{or}{rgb}{0.95,0.65,0}

\begin{document}

\title{A Grand Unified Parity Solution to Strong CP Problem}

\author{Yukihiro Mimura}
\affiliation{Institute of Science and Engineering, Shimane University, Matsue 690-8504, Japan}

\author{ Rabindra N. Mohapatra}
\author{ Matt Severson}
\affiliation{Maryland Center for Fundamental Physics, Department of Physics, 
University of Maryland, College Park, MD 20742, USA}

\begin{abstract} A beyond the standard model theory that respects parity symmetry at short distances is known to provide a solution to the strong CP problem without the need for an axion, while keeping the CKM phase unconstrained. 
	In this paper we present a supersymmetric SO(10) grand unified  embedding of this idea with Yukawa couplings generated by {\bf 10}, ${\bf \overline{126}}$ and {\bf 120} Higgs fields. This model is known to provide a unified description of masses and mixings of quarks  and leptons. When CP symmetry is imposed on this model, the discrete gauge subgroup C of SO(10) combines with it to generate an effective parity symmetry, leading to hermitian quark mass matrices. Imposing an additional discrete symmetry, $G$, we show that there are no other tree level sources of $\theta$ in the model; $G$ also guarantees that the one- and two-loop contributions to $\theta$ vanish.  We then show that the leading three-loop effects and the effect of higher-dimensional operators invariant under $G$ give rise to $\theta$  near the current experimental bound, making the model testable in the current searches for neutron electric dipole moment.
\end{abstract}
\maketitle

One of the major puzzles of the standard model is understanding the smallness of the QCD -induced CP violating parameter $\theta$, which is bounded to be $\leq 10^{-10}$  from the electric dipole moment limits on the neutron. A popular solution to this puzzle is the Peccei-Quinn theory~\cite{PQ} which predicts the existence of a light pseudo-scalar particle, the axion~\cite{WW}, with model dependent mass and couplings. The most widely discussed example is the so-called invisible axion suggested in~\cite{axion}. There have been numerous attempts to experimentally discover the axion, and many new techniques have been proposed recently to search for it. So far such attempts have been unsuccessful.

Several alternative solutions to the $\theta$ problem that do not predict an axion have therefore been discussed in the literature~\cite{hook}. A non-trivial problem here is to make $\theta$ so tiny while keeping the CKM phase so large.
One of the earliest classes of solutions without an axion~\cite{beg, MS} is based on the left-right symmetric theories (LRSM)~\cite{LR}, which extend the standard model to provide a framework for understanding the origin of parity violation. These theories conserve parity ($P$) prior to symmetry breaking, which implies that the quark Yukawa coupling matrices are hermitian if parity transformation $P$ is defined as $Q_L\to Q_R$. If the vacuum expectation values (vevs) of the Higgs fields that break the standard model gauge symmetry are real, the resulting quark mass matrices will be hermitian, leading to $\arg\det M_uM_d\equiv \theta=0$ at tree level while the CKM phase remains unconstrained.

However, since parity symmetry must be broken at a scale higher than $v_{wk}$ to explain observed weak interactions, the question arises as to whether $\theta$ reappears after parity breaking. The answer is that non-zero $\theta$ does appear from quantum loops involving the parity-breaking scale, but the resulting value is finite, calculable, and small. Typically extra symmetries such as supersymmetry are needed~\cite{MR} to suppress $\theta$ to the desired level. For further discussion of such models with supersymmetry, see~\cite{dine}.

The simplest complete parity-based model which does not require any extra symmetries is the quark seesaw version of LRSM~\cite{babu} where $\theta$ arises at the two-loop level and is safely below the current neutron electric dipole moment limits~\cite{babu,hall}. Parity-based solutions that use a $U(1)$-extended version of this model have also been discussed in the literature~\cite{ bcs,hook2,hall2}.  

Another class of widely discussed models  uses spontaneously broken CP invariance~\cite{NB}, but it has been argued~\cite{dine2,hook} that  getting large CKM phase is more of a challenge in these models. For further discussion of these models see~\cite{dine2,luca}.

In this paper, we focus on parity-based models and discuss their embedding into supersymmetric SO(10) grand unified theories. To the best of our knowledge, there has been no grand unified embedding of this kind of solution to the strong CP problem. There are bottom-up models with coupling unification but not full GUT embedding; see~\cite{bdm,hall}. Non-triviality of GUT embedding comes from the fact that there  are many beyond the standard model colored fermions in such models which could potentially contribute an arbitrary amount to $\theta$, spoiling the strong CP solution. Such contributions could arise if, for instance, there are phases in their Yukawa couplings or in vacuum expectation values. This is one of the problems we solve in the present model using discrete symmetries. We then show that the quantum corrections to $\theta$ are also acceptably small.

Our solution uses  a class of recently discussed renormalizable SUSY  SO(10) models with Yukawa-generating Higgs superfields  belonging to {\bf 10}, ${\bf \overline{126}}\oplus {\bf 126}$ and {\bf 120} representations. 
The fact that models of this type have the potential to solve the strong CP problem was noted in~\cite{DMM} where it was pointed out that if CP symmetry is imposed on the model, the quark mass matrices in this model are hermitian. This is because SO(10) contains a C-gauge symmetry as its subgroup, so additionally imposing CP makes the model P-symmetric. Again, in this model, the value of CKM phase is unconstrained. 

The next step was taken in \cite{matt}, where we showed it is possible to construct a superpotential in the model that does not involve any new phases, thus potentially maintaining $\theta^{\rm tree}=0$; it allows for no new phases from the mass matrices of other colored fields in the model. This was done using a $CP\times Z_2$ symmetry. However the full model in~\cite{matt} leads to fields lighter than the GUT scale which interfere with one-step coupling unification.

In this paper we propose a new model with an alternate discrete symmetry that solves the light fields problem while still leading to sufficiently small $\theta$ in the following manner: (i) we provide the superpotential that does not allow any new light states in the theory, maintaining coupling unification as in the MSSM; (ii) We analyze loop effects on the value of $\theta$ and show that a non-zero $\theta$ arises in the form of a gluino mass phase only at the {\it three-loop} level, which contributes a highly suppressed $\theta$, making it a true solution to the strong CP problem. Needless to say the model also provides a fit to all fermion masses including neutrinos and predicts observable proton decay as has been already been discussed in~\cite{DMM,matt}. 

\vskip.1in
\noindent{\underline{\bf The model:}} We work with the SUSY SO(10) theory where Yukawa couplings of SM fermions are generated by Higgs multiplets belonging to  {\bf 10}, ${\bf \overline{126}}$ and {\bf 120}  representations (denoted by $H, \bar\Delta, \Sigma$ respectively). The fermions of the model belong to the {\bf 16} dimensional spinor representation (denoted by $\psi$). The  most general  Yukawa superpotential can be written as
\begin{equation}\label{CP1}
W_Y~=~h_{ij}\psi_i\psi_j H+f_{ij}\psi_i\psi_j\bar{\Delta}+g_{ij}^\prime \frac{Z_\psi}{\Lambda}\psi_i\psi_j\Sigma,
\end{equation}
where $h, f$ are symmetric matrices, $g^\prime$ is an antisymmetric matrix, and 
$Z_\psi$ is a spurion singlet field. 
%
We now require this theory to have an additional CP symmetry under which $\psi \to \psi^*$, $H,\bar{\Delta}\to H^*, \bar{\Delta}^*$, $Z_\psi \to Z^*_\psi$ and $\Sigma\to -\Sigma^*$. Requirement of CP invariance then implies that $h, f$ are symmetric and real matrices and 
$g^\prime$ is an imaginary antisymmetric matrix, 
i.e. $g^{\prime}=ig^{\prime\prime}$ with ${g}^{\prime\prime }$ real. 
We then define $g \equiv  g^{\prime\prime} \langle Z_\psi\rangle/\Lambda$.
%

The MSSM doublets will be linear combination of the ones contained in $H, \bar{\Delta}, \Delta, \Sigma$ and the two up and down type MSSM doublets will be kept at the weak scale by the fine tuning of the Higgs superpotential parameters as is done in usual SUSY GUT theories. After substituting the vevs of the resulting MSSM doublets, we have (see Ref.\cite{DMM} for convention)
\begin{eqnarray}
  {\cal M}_u &=& \tilde{h}+r_2 \tilde{f}+ir_3\tilde{g}, \nonumber \\ 
  {\cal M}_d &=& \frac{r_1}{\tan\beta}
  (\tilde{h}+\tilde{f}+i\tilde{g}), \nonumber \\
  {\cal M}_e &=& \frac{r_1}{\tan\beta}
  (\tilde{h}-3\tilde{f}+ic_e\tilde{g}), \nonumber \\
  {\cal M}_{\nu_D} &=& \tilde{h} - 3 r_2 \tilde{f} + ic_\nu \tilde{g}, \nonumber\\
  {\cal M}_\nu &=& fv_L-{\cal M}_{\nu_D}({fv_R})^{-1}{\cal M}_{\nu_D}^T,
  \label{eq:mass}
\end{eqnarray}
where $\tan \beta = v_u / v_d$ for vevs $v_{u,d}$ of the MSSM
fields $H_{u,d}$. For $\lambda = h,f,g$, the couplings
$\tilde{\lambda}_{ij}$ are related to $\lambda_{ij}$ by~\cite{DMM} 
\begin{equation} 
  \tilde h \equiv {\cal V}_{11} h\, v_u; 
  \  
  \tilde f \equiv \frac{{\cal U}_{14} f v_u}{r_1 \sqrt{3}}; 
  \  
  \tilde g \equiv \frac{{\cal U}_{12} + {\cal U}_{13}/\sqrt{3}}{r_1} g\, v_u.
  \label{CP2}
\end{equation} 
where ${\cal V}, {\cal U}$ are the MSSM Higgs doublet mixing matrices at the GUT scale~\cite{DMM}.

If we can guarantee that the mixings ${\cal U,V}_{ij}$ and vevs $v_{u,d}$ and $\langle Z_\psi\rangle$ are real, the quark and lepton mass matrices will be hermitian, i.e. $M_q~=~M^\dagger_q$, which implies that at tree level, $\theta=\arg\det (M_u M_d)=0$. At the same time, because of $i\tilde{g}$ term in the quark mass matrices, the CKM phase is arbitrary. 

If this model is to be a solution to the strong CP problem, we have to show the following: 
\begin{enumerate}[label=(\roman*)]

\item  the mixing parameters ${\cal U,V}_{ij}$ in MSSM doublets above are real;

\item  the masses of all the GUT-scale colored fermions are real (e.g. $\arg\det M_C= 0$ for colored-Higgs mass matrices $M_C$);

\item there are no higher order loop corrections to $\theta$ that are large; and 

\item  the full superpotential is such that there are no dangerous sub-GUT-scale multiplets that affect coupling unification. 

\end{enumerate}

We show all these below. Clearly the first two require that all the couplings in the Higgs superpotential $\cal W$ are real. We show by an appropriate choice of the discrete symmetry $G$ and choice of $CP$ properties of the superfields that this condition is indeed satisfied in our model.

\vskip.1in
\noindent{\bf \underline{Superpotential} :} In addition to the above multiplets which play a role in generating fermion masses, we add the following multiplets: {\bf 45}, {\bf 54} and {\bf 210} (denoted by $A$, $S$ and $\Phi$ respectively). The CP transformations of the various field in the model are given in the Table 1. In column 3 of the Table, we give the transformation of the fields under a discrete group $G\equiv Z_{N_1}\times Z_{N_2}\times Z_{N_3}$. The purpose of the discrete group is to ensure that there are no new phases in the superpotential that will contribute to the $\theta$ parameter. We have also used some of the Yukawa couplings and masses as spurion fields so that they have appropriate charges under $G$ to make that desired field term $G$-invariant. We will show that the spurions will acquire GUT scale vevs to generate the Yukawa couplings and masses of the right order and discuss how those vevs for the gauge singlet spurion fields arise. 

In Table I, we have assigned the charges to the couplings and masses so that
\begin{itemize}
\item
The Higgs doublets mixed by $H\Sigma A$, $\bar\Delta \Sigma A$ terms
with a vev of $A$,
as well as $S A A$,
are allowed by the renormalizable coupling.

\item
The Yukawa coupling to $\Sigma$ can be suppressed so the charge of $\Sigma$ can be different from
$H$ (we certainly want to write $\psi \psi H$ Yukawa coupling, which generates the top mass, as a renormalizable term).

\item
The masses of $\Delta$, $A$, and $S$, as well as some couplings such as $SHH$,
are treated as spurion singlet fields.


\end{itemize}

\begin{table}\centering
\begin{tabular}{|c||c||c|}\hline
Field & CP transformation & $(Z_{N_1}\times Z_{N_2}\times Z_{N_3})$ charges\\\hline
$\Psi(16)$ & $\Psi^*(16)$ & $(-1/2, 0, 0)$\\
$ H(10)$ & $H^*(10)$& $(1,0,0)$\\
$\bar{\Delta}(\overline{126})$&$\bar{\Delta }{(\overline{126})}^*$ & $(1,0,0)$\\
${\Delta}({126})$ &$\Delta(126)^*$ &$(1,1,0)$\\
$\Sigma(120)$& $-\Sigma(120)^*$& $(1,0,1)$\\
$A(45)$&$-A(45)^*$&$(-2,0,-1)$\\
${S}(54)$&$S(54)^*$&$(4,0,2)$\\
$X_\Delta $,$\Phi(210)$ & $X^*_\Delta$,$\Phi(210)^* $ & $(-2,-1,0)$ \\
$X_\Sigma $ & $X^*_\Sigma $  &$ (-2,0,-2)$ \\
$X_S$ & $X^*_S$&$ (-8,0,-4)$\\
$X_A$&${X^*_A}$&$(4,0,2)$\\
$X_\Phi$&${X^*_\Phi}$&$(4,2,0)$\\
$Z_{ \psi}$ & $Z_{ \psi}^*$& $(0,0,-1)$ \\
$Z_{ H} $& $Z_H^* $& $ (-6,0,-2)$ \\
$Z_{\Phi}$ & $Z^*_{\Phi}$ & $(6,3,0)$\\
$\lambda({\rm Gaugino})$ &  $\lambda^*({\rm Gaugino})$& $(0,0,0)$\\\hline
\end{tabular}
\caption{Charge assignment of the different superfields of the theory. It is understood that complex conjugate superfields have opposite discrete quantum numbers.} \label{table:cp}
\end{table}

The superpotential invariant under $SO(10)\times CP\times G$  (in addition to the Yukawa-like terms in Eq. (1)) is given by:
\begin{eqnarray}
{\cal W}&=& \sum_\varphi X_\varphi \,\varphi^2 + X_\Delta \Delta\bar\Delta+
\lambda_2\, \Sigma\, {A\, H} \nonumber \\
&+&\lambda_4\, \bar{\Delta}\,{A} \,\Sigma 
+ \lambda_5\,{S\,A\,A} +
\lambda_6\, H \Delta \Phi + Z_H\, {S\,H\,H} /\Lambda \nonumber \\ 
&+& \lambda_7\,\Phi\Delta
\bar{\Delta}+\alpha_0 Z_{\Phi} \Phi^3/\Lambda,
\label{eq:cpW}
\end{eqnarray} 
with $\varphi =   \Sigma, {  S}, { A}, \Phi$. 
Note that in this superpotential, all the coupling parameters are real due to CP invariance and the scale $\Lambda$,
 which is assumed to be the string scale,
is much larger than the unification scale $M_U$.
The reality of the couplings implies that all the GUT scale vevs of Higgs fields and the spurions will be real, as will be the mixing parameters ${\cal U,V}_{ij}$. 
Moreover, all the colored Higgs fields will have real mass matrices so that they will not contribute any new phase to the tree level $\theta$ parameter. This establishes our condition (i) and (ii) above
at the leading order in the superpotential.
Higher order non-renormalizable terms can still disturb the conditions (i) and (ii); we will show that their contributions to theta are at or below the current bound.

Additionally, we choose the specific symmetry $G= Z_{24}\times Z_6\times Z_4$ so that we can add the following superpotential terms: 
\begin{eqnarray}\label{spurion-potential}
\!\!\!\!W^\prime&=&\alpha_1\frac{(X_{\Delta}\Delta\bar{\Delta})^2}{\Lambda^3}
+ \alpha_2  \frac{S^6}{\Lambda^3}+\alpha_3 X^3_S+\alpha_4 \frac{Z^4_{\Phi}}{\Lambda}\nonumber\\
&+&\alpha_5\frac{Z_\Phi X^3_\Delta}{\Lambda}
+\alpha_6\frac{X^6_A}{\Lambda^3}+\alpha_7\frac{Z^4_{\psi}}{\Lambda}+
\alpha_8\frac{Z^2_\psi X_AX^2_\Sigma}{\Lambda^2}
\nonumber\\
&+&\alpha_9\frac{Z^3_HX^3_\Sigma}{\Lambda^3}+\alpha_{10}\frac{Z^4_H}{\Lambda}+\alpha_{11}\frac{X^6_\Sigma Z^2_H}{\Lambda^5}
\nonumber\\
&+&\alpha_{12}\frac{X^2_SZ_HX_\Sigma}{\Lambda}+
  \alpha_{13} X^2_AX_S + \alpha_{14} \frac{X_\Phi^3 Z_\Phi^2}{\Lambda^2}\nonumber\\
&+& \alpha_{15} X_\Phi X_\Delta^2  + \alpha_{16} \frac{X_\Phi^6}{\Lambda_3} \nonumber\\
&+&(\mbox{higher order terms in }1/\Lambda).
\end{eqnarray}
With very mild fine tuning of the $\alpha_i$ couplings, the F-term minimization can give 
 GUT scale vevs to all spurion fields as desired. Alternatively spurion vevs could originate 
 from some other mechanisms such as compositeness or from a hidden sector.

\vskip.1in

\noindent{\bf \underline{GUT symmetry breaking and particle spectra:}} 
In this section we show 
by analyzing the superpotential that there are no undesirable light particles below the GUT scale that could destroy coupling unification. 
First note that the higher dimensional terms in Eq.(\ref{spurion-potential}), as well other terms, help to give vevs to all the spurion fields. If we choose the cutoff scale $\Lambda$ to be near the string scale ($\sim 10^{18}$ GeV), by appropriate choice of the coefficients of the higher dimensional terms, we can keep the spurion vevs near the GUT scale ($M_U\sim2\times 10^{16}$ GeV). This will lead to spurion masses below the GUT scale, but being gauge neutral, they do not affect the running of gauge couplings. We have checked that with a mild smallness of the coefficients of the higher dimensional operators, we can keep all the singlet vevs near the GUT scale.

 Next, due to the absence of $A \Delta \bar \Delta$ term,
the F-flatness conditions of $\Delta,\bar \Delta,\Phi$ and $S$,$A$ are separated.
%
%
The $\Delta$ vev, which is at GUT scale along the SU(5)-singlet direction i.e. 
$\langle\Delta_{13579}\rangle=\langle\bar\Delta_{13579}\rangle=v_R\neq 0$ (to get the D-terms to be zero),
can be generated by 
$(\Phi+X_{\Delta})\Delta\bar\Delta+\Phi^3 + X_\Phi \Phi^2$ term.
The vevs of $A$ can be generated by
$X_A A^2 + X_S S^2 + SAA$.
For group theory of such models see~\cite{fuku,aulgarg}.

Note that in the absence of the $\Phi\Delta\bar\Delta$ term,  the superpotential is only a function of the
singlet contraction of $\Delta \bar \Delta$, which implies that the superpotential has a large global symmetry,
and thus the decomposed multiplets under SU(5) are massless.
However, the presence of $\Phi\Delta\bar\Delta$ term cures this problem and makes all submultiplets of $\Delta\oplus\bar\Delta$ massive.
%
\vskip.1in

\noindent{\bf \underline {Solving strong CP}:}

We study the possible generation of the strong CP phase,
by (i) higher order terms, (ii)
loop correction to the gluino mass,
(iii) renormalization contribution to the quark Yukawa coupling as we extrapolate from GUT scale to the weak scale.

\vskip.1in
\noindent{\bf (i) Higher order terms:}

In general one could envision two types of non-renormalizable contribution to $\theta$: (i) one set which are invariant under the discrete symmetry 
$CP\otimes G$ and (ii) terms induced by global symmetry-breaking, non-perturbative gravitational effects. We ignore the latter since there seem to be different opinions on whether black holes really break global symmetries.

%
Given the charge assignments in Table I, we find that the leading operators which can generate phases in the
doublet Higgs mixings and the masses of colored Higgsinos are of dimension-9:
\begin{eqnarray}
&&A \Delta \bar\Delta  X_\Delta Z_{H} Z_{\psi} S^2/\Lambda^5, \nonumber\\
&&H \Sigma \Phi \Delta \bar\Delta Z_{H} Z_{\psi} X_A/\Lambda^5, \nonumber\\
&&A \Phi \Phi X_\Phi Z_H Z_\psi X_A^2/\Lambda^5,
\end{eqnarray}
and the suppression factor will be $(M_U/\Lambda)^5$, where $M_U$ is a GUT unification scale $\sim 2\times 10^{16}$
GeV.
For $\Lambda$ to be the reduced Planck mass, $2.4\times 10^{18}$ GeV, 
the bound for $|\theta| < 10^{-10}$ can be satisfied by these contributions to $\theta$.

It is also the case that a
$\psi \psi A H$ term can break the hermitian nature of the Yukawa coupling matrices,
but such a  term is generated only by a dimension-10 operator under the above charge assignment,
which leads to a $\theta$ below the current bound.

Note that the hierarchy between the cutoff (Planck or string scale) and
the GUT unification is essential to suppress the $\theta$ parameter in the current model. 

\vskip.1in
\noindent{\bf (ii) Loop corrections and gluino phase:} As is well known, phases in all colored fields contribute to the $\theta$ parameter. So we have to track the phases in the gluino mass term in addition to the quark mass matrices and the GUT-scale colored Higgsino mass matrices. At the tree level, gluino mass term is real due to CP symmetry in the Lagrangian. However, CP symmetry is broken  and therefore quantum loops induce non-zero gluino phases. At the one-loop level, the gluino mass is real due to hermiticity of the quark mass matrices, since the contribution from quarks is of the form ${\rm tr}\, {M_q^\dagger A_q}$ where $A_q\propto M_q$ within our setup and all other colored fermion fields have no phases. 

\begin{figure}[htbp]
\center
\includegraphics[clip,width=9cm ]{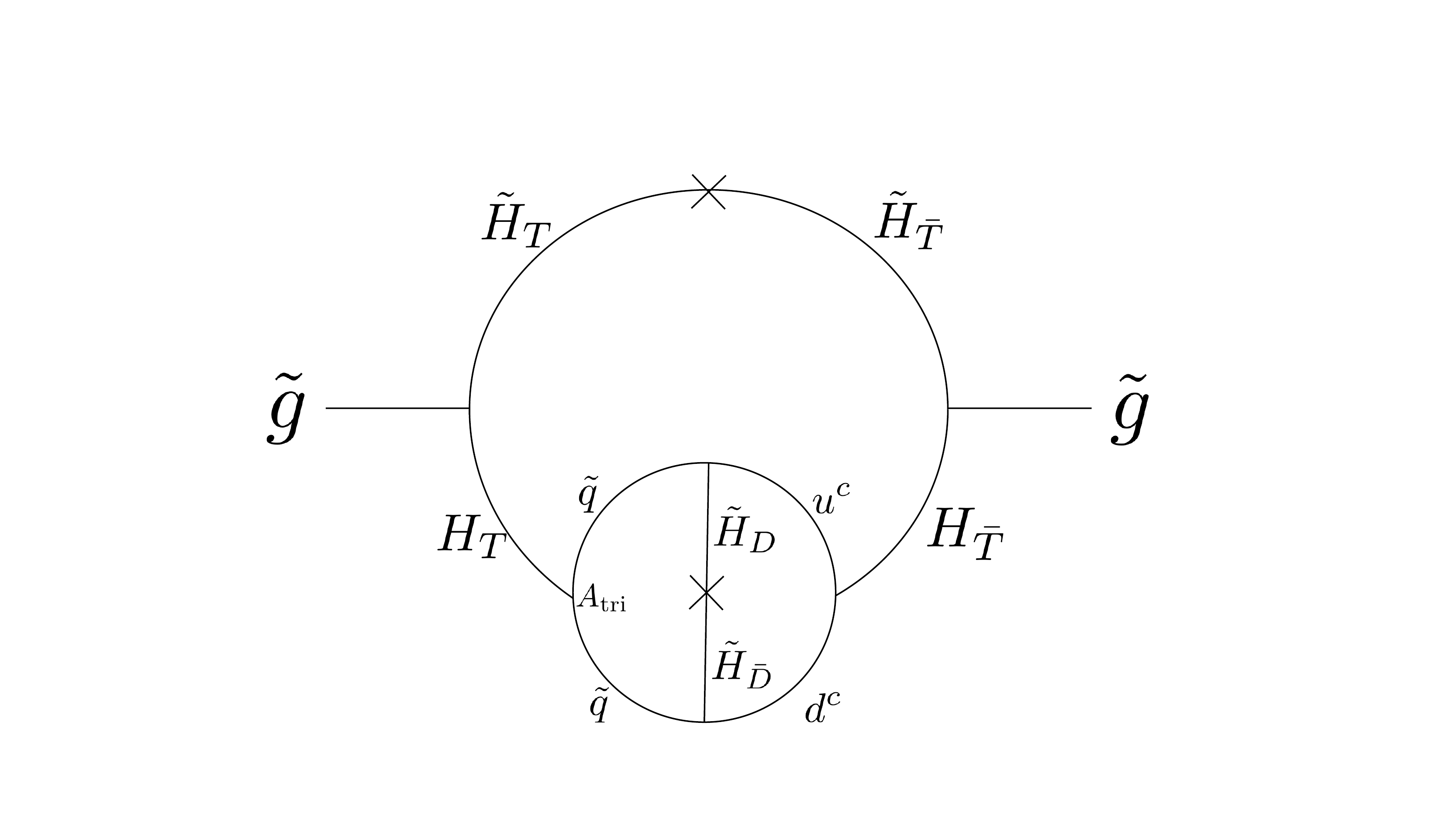}
\caption{The 3-loop diagram which induces the phase of the gluino mass.
There are also diagrams in which $u^c e^c$-$q\ell$ and $\nu^c d^c$-$\ell q$ propagate.
In ${\bf 126}$ and ${\bf 120}$, there are other colored components, e.g. $({\bf 8},{\bf 2},1/2)$,
which have
bi-fermion couplings, and there are similar loop-diagrams in which the other colored components
propagate for the gluino mass correction. }
\end{figure}

The loop correction to the pure-imaginary Yukawa coupling $g$
can induce a phase for the gluino mass
even if all the other couplings and masses are real. 
At the two-loop level,
the contribution is always proportional to ${\rm Tr}\, YY^\dagger$,
where $Y$ is a Yukawa coupling to colored Higgs,
and the contribution is real.
From the three-loop level diagram 
in which the doublet Higgs also propagates (Fig.1),
the gluino mass correction containing the imaginary part is given as
\begin{equation}
\sum_{a,b}\frac{g_s^2}{(16\pi^2)^3}{\rm Tr}\, (Y_{T_a} Y_{D_b}^* Y_{\bar T_a} Y_{\bar D_b}^*) F\left(\frac{M_{H_{D_b}}}{M_{H_{T_b}}}\right) A_{\rm tri},
\end{equation}
where 
$Y_T$ and $Y_{\bar T}$ are Yukawa couplings to (diagonalized) colored Higgs fields ($H_{T,\bar{T}}$),
and $Y_D$ and $Y_{\bar D}$ are the Yukawa couplings to doublet Higgs fields ($H_{D,\bar{D}}$),
and $A_{\rm tri}$ is the SUSY breaking scalar trilinear coupling,
and $F$ is a loop function.
(We note that the heavy Higgs doublets and all the Higgs triplets (not only the lightest ones)
around the GUT scale propagate in the loop diagram.)
The Yukawa couplings are given by the linear combination of
$h$, $f$ and $g$.
Noting that ${\rm Tr}\, (g X) = 0$ for symmetric matrix $X$,
one finds that ${\rm Tr}\, (g h^3)$, ${\rm Tr}\, (g h f h)$, etc vanish,
and ${\rm Tr}\, (g hhf)$ and ${\rm Tr}\, (g ff h)$ etc can contribute\footnote{
It is interesting to note 
that the existence of two different symmetric matrices, $h$ and $f$, 
can induce the phase of the gluino mass.
We also note that this contribution vanishes
near the $SU(2)_L\times SU(2)_R\times SU(4)_c$ vacua where due to
 $SU(2)_R$ symmetry (i.e., ${\rm Re} Y_D = {\rm Re} Y_{\bar D}$),
 the leading contribution for $q q$-$u^c d^c$ sub-diagram in Fig. 1 vanishes.
}.
As a result,
the leading contribution can be estimated
as
\begin{equation}
{\rm Im}\, m_{\tilde g} \sim \frac{g_s^2 g_{23} f_{23} h_{33}^2}{(16\pi^2)^3} A_{\rm tri}.
\end{equation}
For $A_{\rm tri}\sim m_{\tilde g}$,
we roughly estimate 
the contribution to $\theta$ as
\begin{equation}
\Delta  \theta \sim 10^{-9}  \left|\frac{f_{23} g_{23}}{10^{-3}}\right|.
\end{equation}
This is on the borderline of satisfying the neutron EDM bounds,
taking into account that $f$ and $g$ are the original couplings and 
not multiplied by the Higgs mixing.\footnote{
$\tilde f_{23}$ and $\tilde g_{23}$ (which are $f$ and $g$ multiplied by Higgs mixings)
can be estimated to be as large as $V_{cb}$.  }
If $A_{\rm tri} \ll m_{\tilde g}$, the neutron EDM bounds can be safely satisfied;
in this sense, gauge-mediated SUSY breaking, rather than the gravity mediation, is preferable
to suppress the $\theta$ parameter.
In any case, loop correction to the gluino mass can generate a borderline value for the
$\theta$ parameter, and therefore, the model would predict an observable
neutron EDM in current experiments. 

\vskip.1in
\noindent{\bf (iii) Extrapolation from GUT scale to weak scale for $\boldsymbol\theta$:} The hermiticity of the quark mass matrices holds at the GUT scale. This means that the value of tree level $\theta$ is zero at that scale and a finite, non-zero $\theta$ will be induced at the weak scale due to renormalization extrapolation of the various Yukawa. This issue for a parity solution to strong CP has been considered in~\cite{bdm}, and it is shown that if the weak scale theory is MSSM, the corrections to $\theta$ are given by
\begin{eqnarray}
\delta\theta\simeq 
\left(\frac{1}{16\pi^2}\ln \frac{M_U}{M_W}\right)^4\left[c_1 {\rm Im} {\rm Tr}[Y^2_uY^4_dY^4_uY^2_d]\right.\nonumber\\+c_2 (u\leftrightarrow d)].
\end{eqnarray}
This can be estimated to be $\delta\theta \sim 10^{-26}\,(\tan\beta)^6(c_1-c_2)$, which is below the experimental upper limit even for $\tan \beta=50$.
We note that one obtains $c_1=c_2$ and $\delta\theta$ vanishes without an electroweak gauge loop.

Finally, we note that since the discrete symmetries of our model break at the GUT scale, domain walls associated with them will get ``inflated away" as long as the reheat temperature is below the GUT scale and will not lead to any anisotropy in cosmic microwave background.

\vskip.1in
\noindent{ \bf Acknowledgements:} We thank Anson Hook for helpful discussions and comments on the manuscript.
The work of Y.M. is supported by Scientific Grants by the Ministry
of Education, Culture, Sports, Science and Technology of Japan
(Nos.~16H00871, 16H02189, 17K05415 and 18H04590) and that of R. N. M. is supported by the US National Science Foundation under Grant No. PHY1620074.

\end{document}